\documentclass[fleqn,usenatbib]{mnras}

\usepackage{newtxtext,newtxmath}

\usepackage[T1]{fontenc}
\usepackage{ae,aecompl}

\usepackage{graphicx}	
\usepackage{amsmath}	
\usepackage{amssymb}	

\usepackage{color}

\title[Neutral beam heating of ISM]{Does Nature use neutral beams for interstellar plasma heating around compact objects?}

\author[E. Churazov et al.]{
E. Churazov,$^{1,2}$\thanks{E-mail: churazov@mpa-garching.mpg.de}
I. Khabibullin,$^{1,2}$
R. Sunyaev$^{1,2}$
\\
$^1$~Max Planck Institute for Astrophysics, Karl-Schwarzschild-Str. 1, D-85741 Garching, Germany  \\
$^2$~Space Research Institute (IKI), Profsoyuznaya 84/32, Moscow 117997, Russia 
}

\date{Accepted XXX. Received YYY; in original form ZZZ}

\begin{document}
\label{firstpage}
\pagerange{\pageref{firstpage}--\pageref{lastpage}}
\maketitle

\begin{abstract}
A neutral beam injection technique is employed in all major TOKAMAK facilities for heating of magnetically confined plasma. The question then arises, whether a similar mechanism might work in astrophysical objects? For instance, a hyper-Eddington galactic binary SS433 possesses baryonic jets, moving at a quarter of the speed of light, and observations revealed signs of gas cooling and recombination on sub-pc scales and equally strong signs of powerful energy deposition on much larger scales $\sim$100 pc. Here we consider a model where neutral atoms transport this energy. A sub-relativistic beam of neutral atoms penetrates the interstellar medium, these atoms gradually get ionised and deposit their energy over a region, whose longitudinal dimension is set by the "ionisation length". The channel, where the energy is deposited, expands sideways and drives a shock in the lateral direction. Once the density in the channel drops, the heating rate by the beam drops accordingly, and the region of the energy release moves along the direction of the beam. We discuss distinct features associated with this scenario and speculate that such configuration might also boost shock acceleration of the "pick-up" protons that arise due to ionisation of neutral atoms both upstream and downstream of the shock.  
\end{abstract}




\section{Introduction}
Compact astrophysical objects -- neutron stars or stellar-mass/supermassive black holes -- are known to launch energetic outflows, powered by their rotation, magnetic fields and the gravitational energy liberated in the process of matter accretion. These outflows are capable of transporting an immense amount of energy and momentum to large distances away from the source, eventually depositing it into the ambient diffuse medium. Despite being crucially important for several key galactic- and intergalactic-scale phenomena, such as the energy content in the turbulent multiphase ISM or SMBH feedback, the underlying physical processes behind the interaction of the outflow with the gas remain unclear.

While in the diffuse astrophysical plasma the mean-free-path for Coulomb scattering is often large, especially for fast particles, their Larmor radii are many orders of magnitude smaller. Strong coupling of the charged particles through the magnetic field permeating the plasma allows, to the first approximation, treating the outflow/gas interaction in the frame of conventional (magneto)hydrodynamics. More elaborate models consider various plasma effects, allowing for anisotropy of particle distribution functions and associated transport processes. 

The behaviour of neutral particles residing in an outflow (if any) can be very different. Indeed, neutral particles can traverse freely through the magnetised plasma with their mean-free-path being set by atomic processes, e.g., ionisation, charge exchange or elastic scattering. Such a mean-free-path can be large, in some cases comparable with the characteristic system size, and, of course, much larger than the Larmor radius of charged particles. Once ionized, these particles start gyrating in the magnetic field and from now on they are strongly coupled to the plasma. This property of neutral particles is employed in major TOKAMAK facilities to heat the plasma isolated from the walls by magnetic fields using a technique known as Neutral Beam Injection (NBI)  \cite[see, e.g.][]{2009NucFu..49d5006H}. A variant of this technique is also known as a "wet wood burner" concept \citep[e.g.][]{Dean1998}.  Fast neutral particles are also observed in space plasmas. For instance, Energetic Neutral {\bf Atoms} (ENA) exist in the solar system, where they form by via  charge exchange of fast ions with neutral (slow) atoms \cite[see, e.g.][]{1997RScI...68.3617G}. Charge exchange also gives rise to the so-called "pick-up" ions in the solar wind \cite[e.g.][]{1987JGR....92.1067I}. In all these cases, the coexistence of neutral and ionised phases having different velocities leads to a number of nontrivial phenomena.     

Hyper-accreting X-ray binaries such as, e.g. SS 433 are known to possess powerful baryonic jets \cite[see, e.g.][for reviews]{1984ARA&A..22..507M,2004ASPRv..12....1F} moving with  sub-relativistic velocities relating to the ambient ISM. The jets in SS 433 are characterized by progressive cooling through the X-ray regime followed by intense $H_\alpha$ emission some $10^{15}\,{\rm cm}$ from the central source. In addition to radiative cooling losses, the adiabatic expansion is certainly contributing to the cooling of the outflowing gas.  While the details of the SS 433 jets behaviour at larger distances are not well understood, it is conceivable that for hyper-accreting objects the interaction of a fast beam of neutral particles with the ISM might play a major role. Here we consider the most basic implications of such setup. Under certain conditions, the presented picture might be relevant for various situations, where highly supercritical accretion is believed to take place: e.g. ultraluminous X-ray sources in nearby galaxies, tidal disruption events and growing seeds of supermassive black holes in the early Universe.

\section{Basic model}

\begin{figure}
\centering
\includegraphics[trim= 0mm 0cm 0mm 0cm,width=1.\columnwidth]{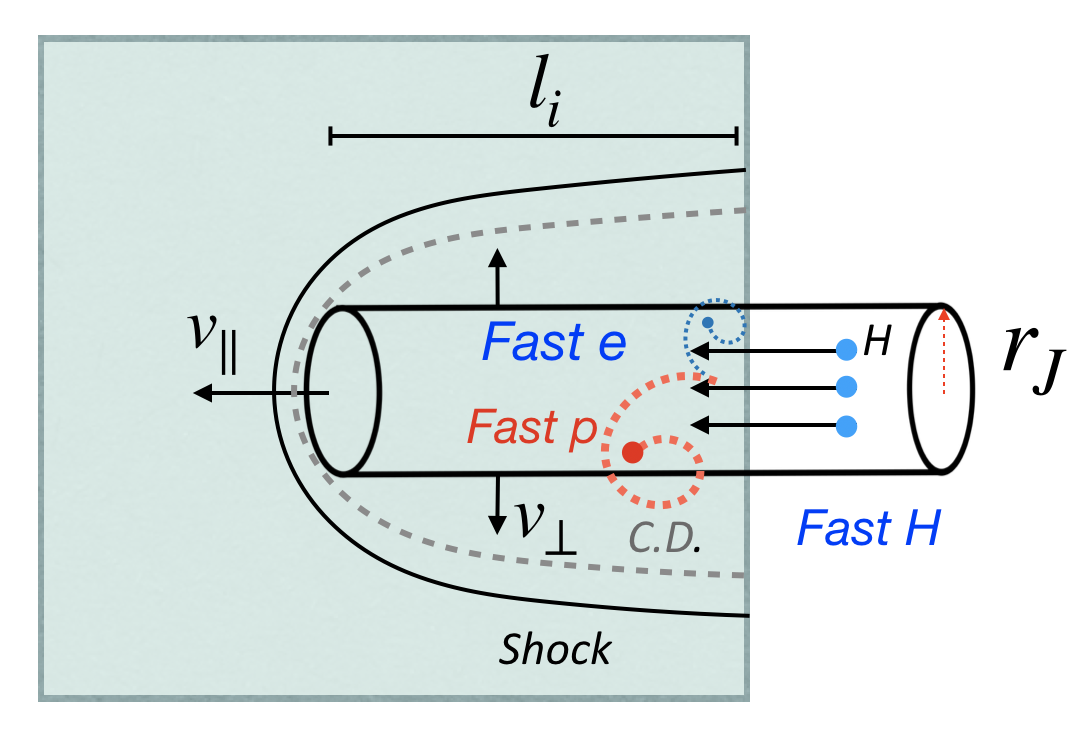}
\caption{Geometry of the problem: a cylindrical beam of fast neutral hydrogen atoms enters the ISM (gray region). The atoms gets ionized over a distance $l_i=\displaystyle (\sigma_i\times n_{\rm ISM})^{-1}$. Energetic electrons and protons arising from the ionisation then couple with the magnetised ISM, sharing their energy and momentum with the ambient gas. The increased pressure in the deposition region forces it to expand, driving a shock in the surrounding medium and decreasing its density. The drop in density allows neutral particles in the beam to propagate further along the direction of the beam, where they will be ionized and deposit their energy in the yet unperturbed volume of the ISM. See text for estimates of the lateral expansion and the beam propagation velocities  $v_\perp$ and $v_\parallel$, respectively. The dashed gray line show schematically the contact discontinuity (C.D.), separating the beam-heated gas from the shocked ISM.}
\label{fig:geometry}
\end{figure}

Consider a cylindrical beam of fast hydrogen atoms moving with the velocity $v=\beta c$, where $\beta\sim 0.26$ (as observed in SS 433) and $c$ is the speed of light. At such velocity, the kinetic energy is $\sim 32$~MeV and $\sim 17$~keV for a proton and electron, respectively. The beam is assumed to be kinetic, i.e. collisions between individual hydrogen atoms within the beam are very rare. The beam enters a region filled with a low-temperature ($T\sim 10^4$~K)  and possibly partially ionized gas with density $n_{ISM}$, which is much higher than the beam density (see Fig.~\ref{fig:geometry}). 

At these energies (velocities), interaction between fast hydrogen atoms and particles of the medium proceed mostly via ionization, since charge exchange cross section is much smaller and we will neglect its contribution in the rest of the paper. The ionization cross-section of the fast atoms by electrons, protons and neutral hydrogen atoms of the medium is \citep[e.g.][]{1960PPS....75..374D,1968ZPhy..216..241L}
\begin{equation}
\sigma_i\sim 3\times10^{-19}-10^{-18}\, {\rm cm^{2}}\, \left ( \frac{E_p}{30~{\rm MeV}} \right )^{-1},
\label{eq:sigmai}
\end{equation}
where $E_p=m_{p}v^2/2=m_p c^2 \beta^2/2$ is the atoms's kinetic energy in the non-relativistic regime (which is of primary interest here).


Accordingly, these fast hydrogen atoms will be ionized after moving over a characteristic length
\begin{equation}
l_i=\frac{1}{n_{ISM}\sigma_i}\approx 1\,{\rm pc} \, \left(\frac{n_{ISM}}{1\, {\rm cm}^{-3}}\right)^{-1}\left(\frac{\sigma_{i}}{3\times10^{-19}\, {\rm cm}^{2}}\right)^{-1} 
\end{equation}

Once the fast hydrogen atom is ionized, the resulting fast proton and electron should rapidly couple to the medium via magnetic field and contribute to the local energy density. As a result, the neutral beam effectively deposits all its energy flux (i.e. kinetic luminosity) $L_J$ (${\rm erg\,s^{-1}}$) in the region of length $l_{i}$. Given the high velocity and low density of the beam, the momentum and mass deposited by it might be neglected, at least in the first approximation. 

The heating rate per unit volume of the medium exposed to the cylindrical neutral beam of radius $r_J$ (see Fig.\ref{fig:geometry}) equals
\begin{equation}
\dot{\epsilon}=L_{J}\,n_{ISM}\,\frac{\sigma_i}{\pi r_J^2}.
\end{equation}
The beam-crossing-time of this region
\begin{equation}
t_i={l_i}/{v}\approx 12\,{\rm yr} \, \left(\frac{n_{ISM}}{1\, {\rm cm}^{-3}}\right)^{-1}\left(\frac{\sigma_{i}}{3\times10^{-19}\, {\rm cm}^{2}}\right)^{-1}\left(\frac{v}{0.26c}\right)^{-1}  
\end{equation}
is clearly sufficiently short for the injection to be treated as instantaneous across this region.

Hereafter, we will assume the following fiducial parameters: $L_J=10^{39}\,{\rm erg\,s^{-1}}$, $r_J=3\,{\rm pc}$, $n_{ISM}=1\,{\rm cm^{-3}}$, $\sigma_i=3\,10^{-19}\,{\rm cm^2}$.  It turns out that for such parameters, the heating rate is high and the strongly over-heated volume should start expanding, sending a shock into the surrounding undisturbed medium.

Assuming that $l_i \gg r_J$, 
 we can estimate (omitting several factors of order unity) the "lateral" expansion velocity $v_\perp$ of the heated region and its expansion time $t_e$, by requiring two conditions
\begin{eqnarray}
(n_{ISM}\, m_p\, l_i\, \pi r_J^2) v_\perp^2 \sim L_J\,n_{ISM}\,l_i\,\sigma_i\, t_e \left [= L_J\, t_e \right ]\\
\frac{r_J}{v_\perp} \sim t_e, 
\end{eqnarray}
where the first condition equates the kinetic energy of the expanding gas to the total deposited energy, while the second condition evaluates the $r_J$ crossing time. Effectively, $t_e$ is the time interval during which the expansion of the heated region can be neglected. The above conditions yield the following estimates
\begin{eqnarray}
t_e\sim 1600~{\rm yr} \left(\frac{r_J}{3\,{\rm pc}} \right)^{4/3} \left(\frac{L_J}{10^{39}\,{\rm erg\,s^{-1}}} \right)^{1/3} \left(\frac{\sigma_i}{3\,10^{-19}\,{\rm cm^2}} \right)^{-1/3}
\end{eqnarray}
and
\begin{equation}
v_\perp\sim 
1800\,{\rm km\,s^{-1}}\left ( \frac{L_{J}}{10^{39}\,{\rm erg\,s^{-1}}}\right)^{-1/3}\left ( \frac{r_{J}}{3\,{\rm pc}}\right)^{-1/3} \left(\frac{\sigma_i}{3\,10^{-19}\,{\rm cm^2}} \right)^{1/3}.
\label{eq:v_perp}
\end{equation}
Notice, that neither $t_e$ nor $v_\perp$ depend on the ISM density. 

Of course, the above estimates are valid as long as the heated region does not expand. Once $t\gtrsim t_e$, the lateral expansion reduces the density of the gas in the volume exposed to the beam. Therefore, (i) the total heating rate in this volume goes down and (ii) due to reduced density the beam is now able to propagate further into the ISM.  The corresponding propagation speed is set by the condition 
\begin{equation}
v_\parallel \sim \frac{l_i}{t_e}\sim 
600\,{\rm km\,s^{-1}}\left ( \frac{n_{ISM}}{1\,{\rm cm^{-3}}}\right)^{-1}\left ( \frac{L_{J}}{10^{39}\,{\rm erg\,s^{-1}}}\right)^{1/3}\left ( \frac{r_{J}}{3\,{\rm pc}}\right)^{-4/3}.
\label{eq:v_parallel}
\end{equation}
The resulting velocity is much smaller than the assumed beam velocity $0.26\,c\approx 80\,000\,{\rm km\,s^{-1}}$. In this limit, actual beam velocity enters only via the dependence of $\sigma_i$ on the energy of the colliding particles (see Eq.~(\ref{eq:sigmai})). 

Thus, the overall picture can be summarized as follows: at any given moment the beam is depositing much of its energy over a distance $\sim l_i$. After the time $\sim t_e$, the lateral expansion reduces the local gas density (and, therefore, the local heating rate) and the region of the main energy release propagates further, leaving behind a blast wave that expands sideways\footnote{We note in passing that even after expansion of the heated volume, the beam continues to pump energy into this region, albeit at a reduced rate.}.  If the length of the affected region is much larger than $l_i$, this means that the beam was operating long enough, so that the gas in many $l_i$-long portions has been sequentially heated and expanded. 
A comparison of Eqs.~(\ref{eq:v_perp}) and (\ref{eq:v_parallel}) shows that the ratio of  $v_\perp$ to $v_\parallel$ 
\begin{equation}
\frac{v_\perp}{v_\parallel}\propto (n_{ISM} \, \sigma_i \, r_J) \, L_J^{-2/3} = \frac{r_J}{l_i} \, L_J^{-2/3}
\end{equation}
directly depends on the ratio of $r_J$ to $l_i$. This factor is expected to affect the aspect ratio of the forming structures.

\section{Illustrative numerical examples}
\label{sec:num}
In order to illustrate the above arguments, we have performed a set of simple hydrodynamic simulations using publicly available hydro code \verb# PLUTO # \citep{2007ApJS..170..228M} under assumption of cylindrical symmetry. The grid is uniform with reflective boundary conditions on each side of the computational domain. The $X$ axis is along the direction of the beam (increasing leftwards), while the $Y$ axis is along the radius of the beam. A cylindrical beam is moving from the right to the left and is modelled as a distributed energy release due to gradual ionisation of hydrogen atoms in the following form: 
\begin{equation}
\dot{\epsilon}(x,y,t)=L_{J}\,n\,\frac{\sigma_i}{\pi r_J^2}\times e^{-\tau(x,y,t)}{\rm rect}\left(\frac{y}{2\,r_j}\right),
\end{equation}
where $\displaystyle \tau(x,y,t)=\int_{0}^{x} n(\zeta,y,t) \sigma_i d\zeta$ is the column density of the gas along the trajectory of neutral atoms and $n(\zeta,y,t)$ is the density of the ISM. Initially, $n(\zeta,y,t=0)={\rm const}=n_{ISM}$, but it evolves with time due to expansion. Various subtle effects, for instance, the change of the ISM ionization state are ignored. We also ignore the contribution of the beam to the momentum and density of the fluid after neutral atoms are ionized.  With these assumptions, the neutral beam acts merely as a distributed source of heat.  For the ISM density in the range $10^{-2}-1\,{\rm cm^{-3}}$, the characteristic (radiative) cooling time $t_{cool}$ is longer than $10^5$~yr, which is in turn much longer than the other time scales of interest here. Therefore, the radiative losses can be neglected.

\begin{figure}
\centering
\includegraphics[angle=0,trim= 0mm 5cm 0mm 5.6cm,clip,width=0.95\columnwidth]{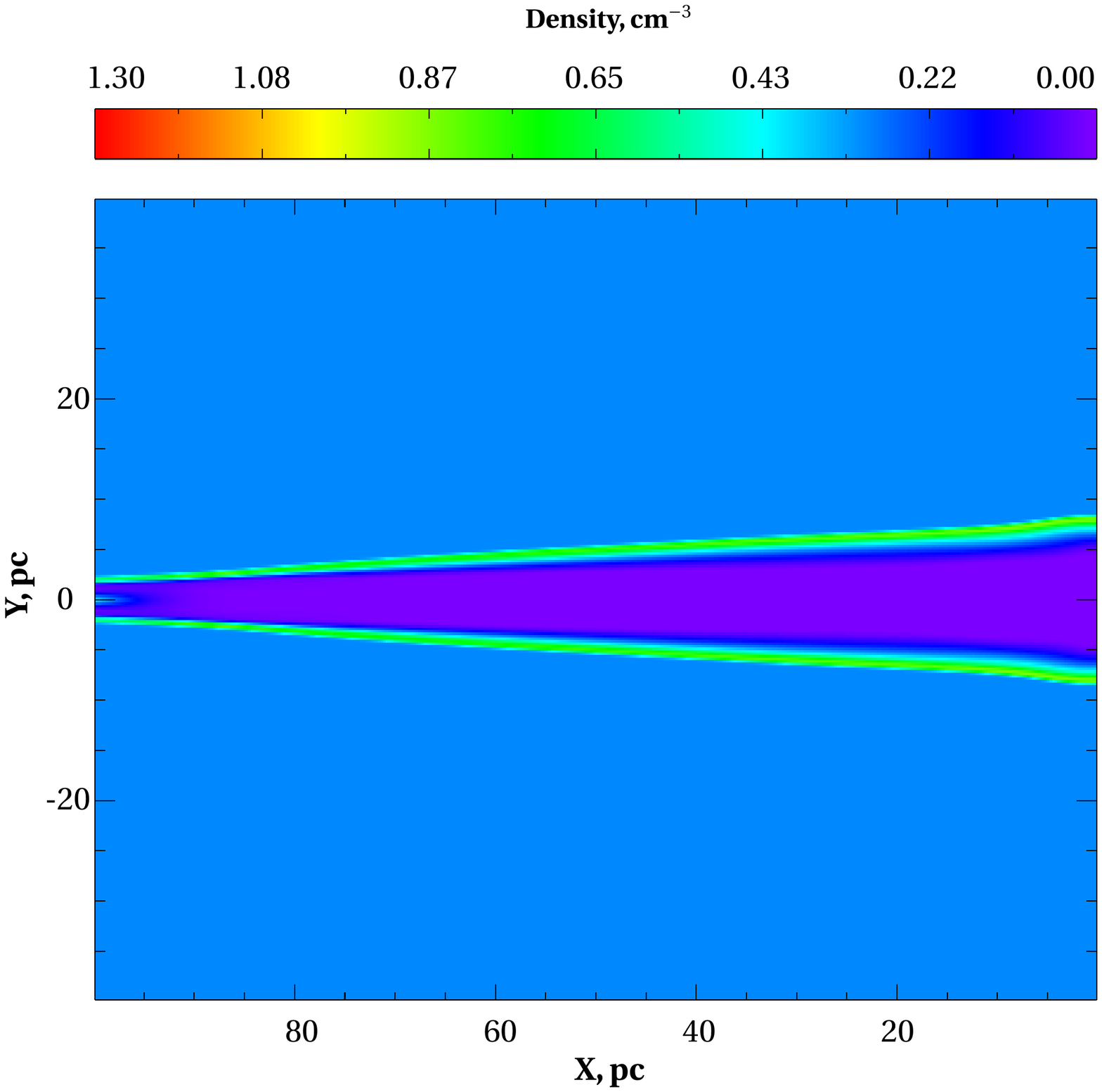}
\includegraphics[angle=0,trim= 0mm 5cm 0mm 8.5cm,clip,width=0.95\columnwidth]{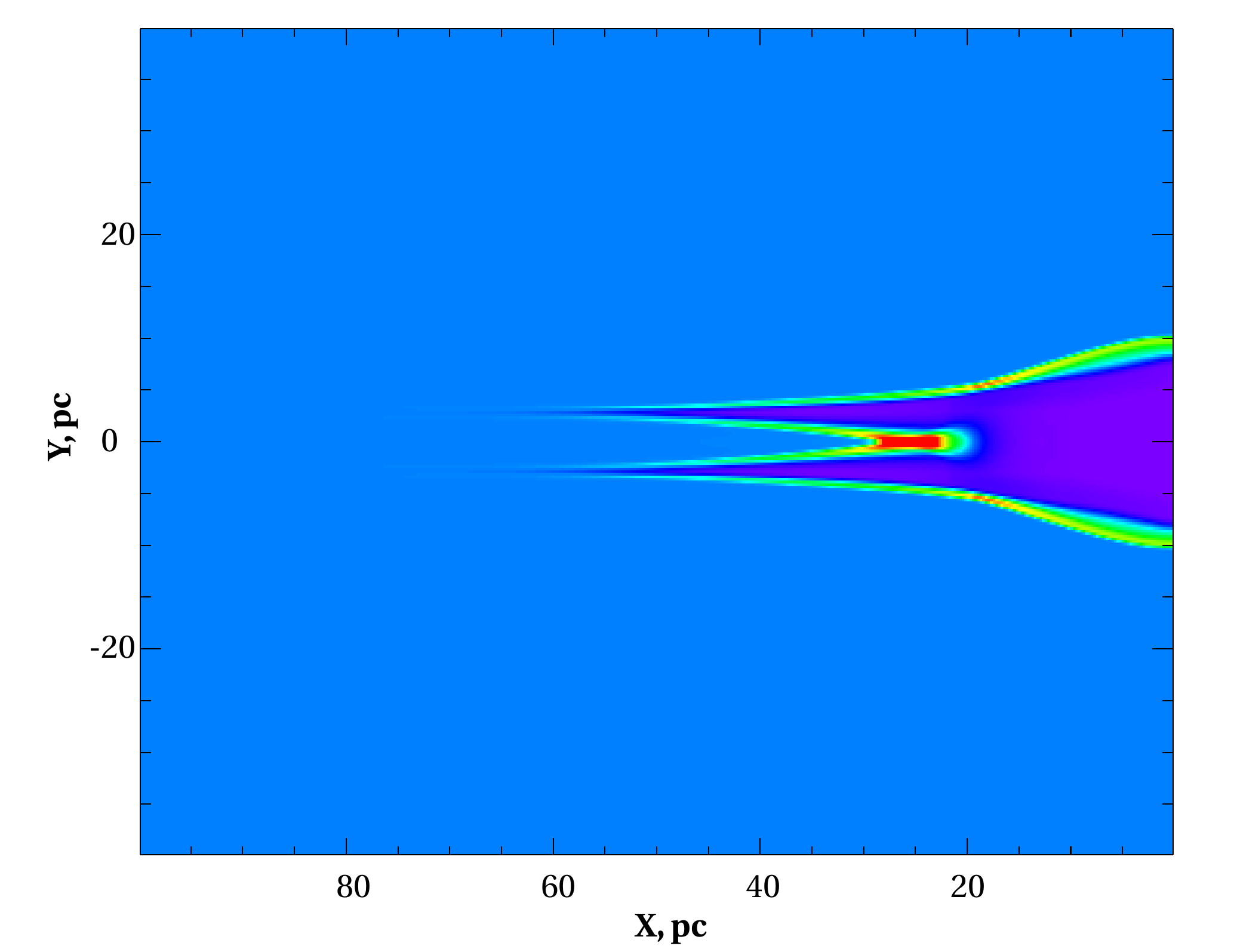}
\includegraphics[angle=0,trim= 0mm 5cm 0mm 8.5cm,clip,width=0.95\columnwidth]{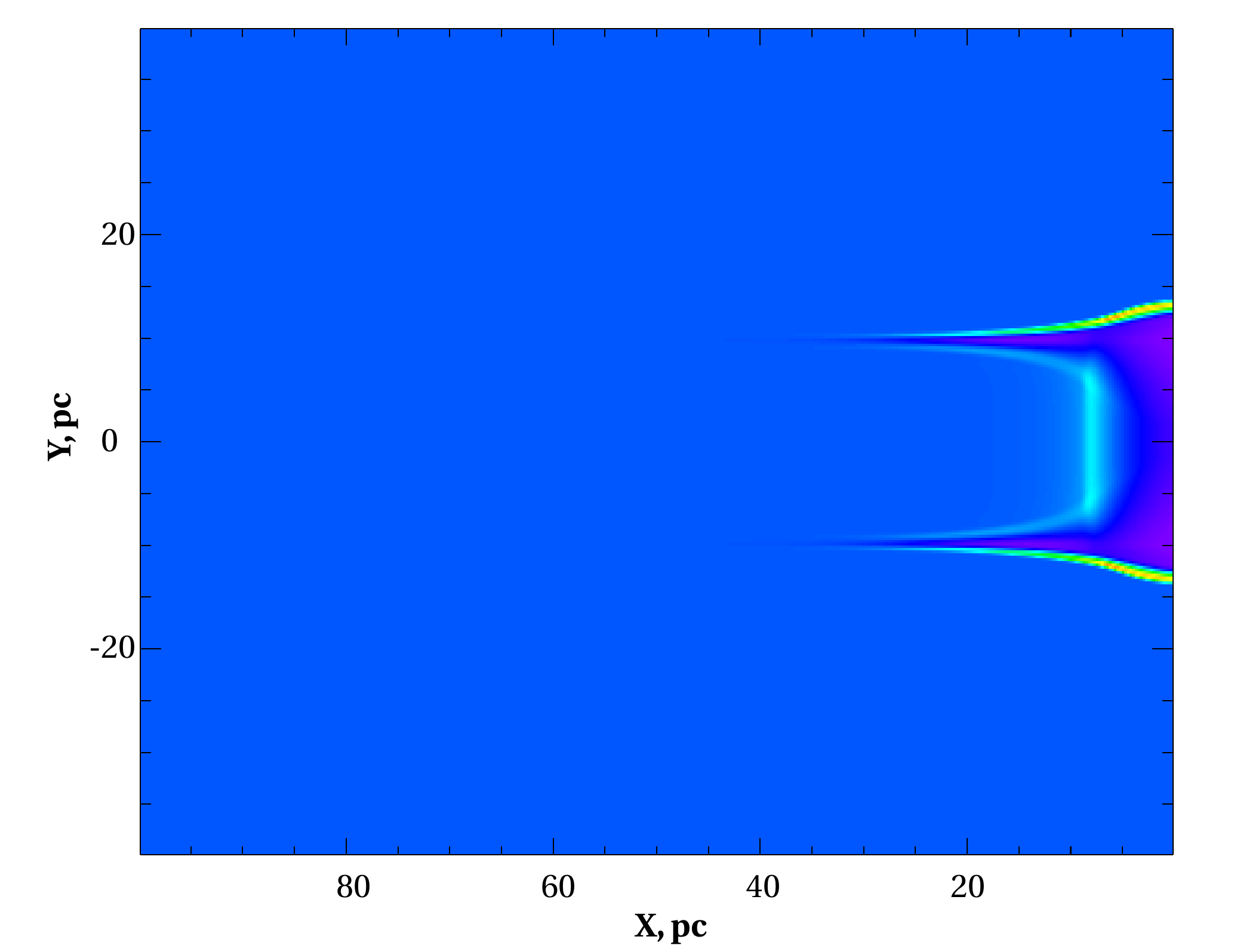}
\caption{Simulations of a neutral beam in the ISM. A snapshot of the gas density $10^4$~yr after the beginning of simulations is shown. In all three frames, the same values of $L_J=10^{39}\,{\rm erg\,s^{-1}}$,  $n_{\rm ISM}=0.3\,{\rm cm^{-3}}$ and $\sigma_i=3\,10^{-19}\,{\rm cm^2}$ are used, while $r_J$ is set to 1, 3 and 10~pc for top, middle and bottom panels, respectively. Three important features to notice in these plots are: (i) as expected (Eq.~\ref{eq:v_parallel}), a narrow beam (top panel) propagates faster than wide beams, (ii) the head of the heated region has a "double-horn" structure in the 2D slices (i.e. a surface of a cylinder in 3D), and (iii) a dense (and hot) region is formed by a converging cylindrical shock, best seen in the middle panel (see \S\ref{sec:num}).
\label{fig:nb_rj}
}
\end{figure}

Fig.~\ref{fig:nb_rj} shows the density distribution after $10^{4}$~yr since the beginning of the simulations. The initial density and the power of the beam are the same for all three panels, while the radius of the beam $r_J$ is set to 1, 3 and 10~pc for the top, middle and bottom panels, respectively. As expected (see Eqs.~(\ref{eq:v_perp}) and (\ref{eq:v_parallel})), the narrow beam leads to faster expansion of the heated channel and, as a result, faster propagation of the beam. 

More odd is a double-horn structure, which is present in the all three simulations and most clearly visible in the runs with a large $r_J$. In 3D, this structure corresponds to a "tube" along the surface of the beam. This structure forms naturally since the expansion of a uniformly-heated cylindrical region starts from its outer boundary, while the regions closer to the axis will be involved in expansion after some delay. Thus, the ISM density will first drop in the vicinity of the beam "walls", where neutral atoms can now propagate much further along the beam direction than in the central (still dense) regions of the beam. Therefore, a "hot tube" is formed ahead of the main beam, which expands and sends a pair of shocks, one moving away from the beam and another one moving towards the axis of the beam. 

A by-product of this behaviour is a dense and hot region that appears when the inward moving shocks collide near the axis of the cylinder. This region is the most clearly seen in the middle panel of Fig.~\ref{fig:nb_rj} (red elongated area at $x\sim 30$ and $y\sim 0$). This high-density region ("dense core" hereafter)  moves together with the front part of the beam. We note here, that the formation of the "hot tube" and the "dense core" depends principally on the distribution of beam power in the direction perpendicular to the beam axis. For instance, for a Gaussian distribution, these structures can be much less prominent or even absent, depending on the combination of the parameters. The rule of thumb is that if the distribution is boxy, then these structures are more likely to form.    

\begin{figure}
\centering
\includegraphics[angle=0,trim= 0mm 4.5cm 0mm 5.5cm,clip,width=0.95\columnwidth]{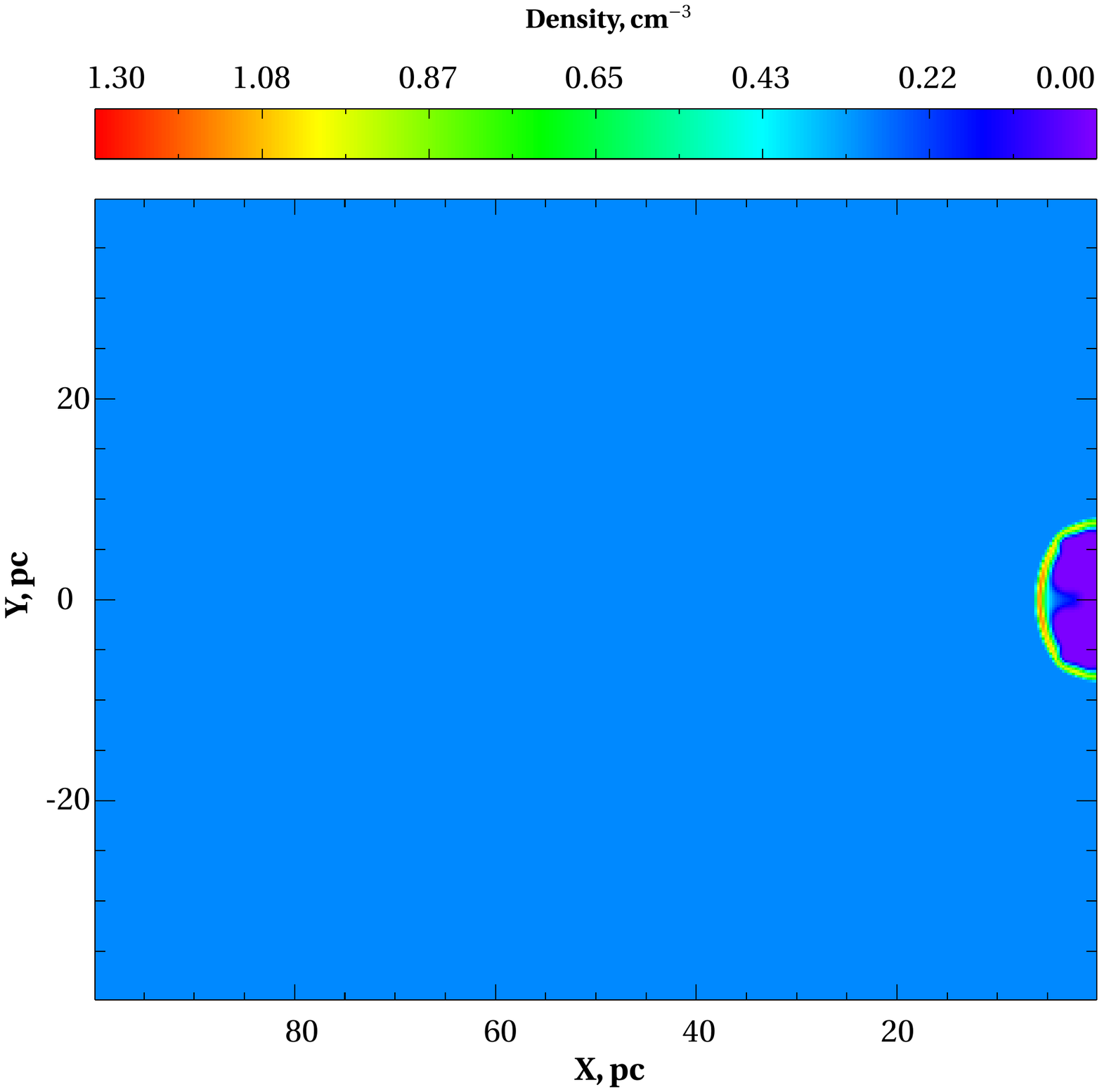}
\includegraphics[angle=0,trim= 0mm 5cm 0mm 5.5cm,clip,width=0.95\columnwidth]{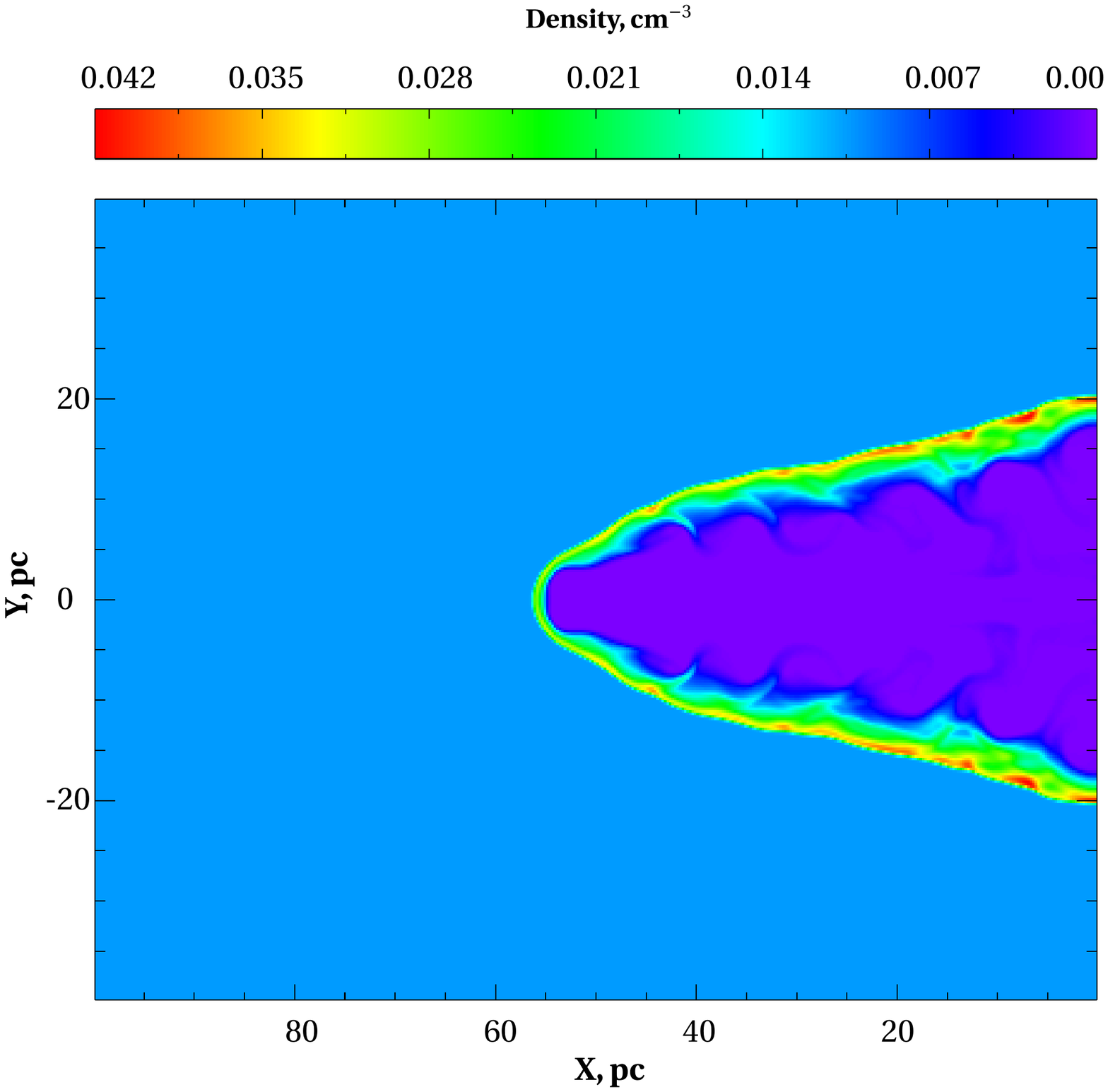}
\caption{Simulations of a hydrodynamic jet propagation in the ISM. In the {\bf top panel}, the parameters of the jet and the ISM are same as in the middle panel of Fig.~\ref{fig:nb_rj}. It is clear that the hydrodynamical jet propagates much slower than the neutral beam. The  bottom panel shows the simulations with a 30 times lower ISM density and 3 times narrower jet. With such parameters, the propagation velocity becomes comparable to the neutral beam run, shown  middle panel of Fig.~\ref{fig:nb_rj}. }
\label{fig:hb_def}
\end{figure}

It is interesting to compare the impact of a neutral beam (hereafter NB) on the ISM with that of a hydrodynamic jet having the same density and velocity. The top panel in Fig.~\ref{fig:hb_def} shows the results of a pure hydrodynamic run, where all parameters ($n_{ISM},r_J,L_J,\beta$) are the same as for the default neutral beam case (see middle panel in Fig.~\ref{fig:nb_rj}), but the beam is now treated as a fluid. One can see that the hydrodynamic jet (hereafter HJ) propagates much slower than the neutral beam. Indeed, the velocity of the jet head propagation is set by the condition \citep{1974MNRAS.169..395B,1984RvMP...56..255B} 
\begin{eqnarray}
v_{\parallel,HD}\sim v_j\sqrt{\frac{n_J}{n_{ISM}}}\propto L_J^{1/2}\,r_J^{-1}\,v_J^{-1/2}\,n_{ISM}^{-1/2}, 
\end{eqnarray}
which is different from eq.~(\ref{eq:v_parallel}). For the HJ case, the momentum of the fluid plays an important role, which is low for fast jets. One can make the propagation of a HJ faster by lowering the density of the ISM and/or making the jet narrow. An example is shown in the lower panel of Fig.~\ref{fig:hb_def}, where the ISM density is 30 times lower than in the fiducial case, while the jet is 3 times narrower (the total jet energy flux is kept unchanged). With this choice of parameters, the propagation velocity of the HJ is comparable to that of a wider NB in a much denser environment.

\begin{figure}
\centering
\includegraphics[angle=0,trim= 0mm 5cm 0mm 5.5cm,clip,width=0.95\columnwidth]{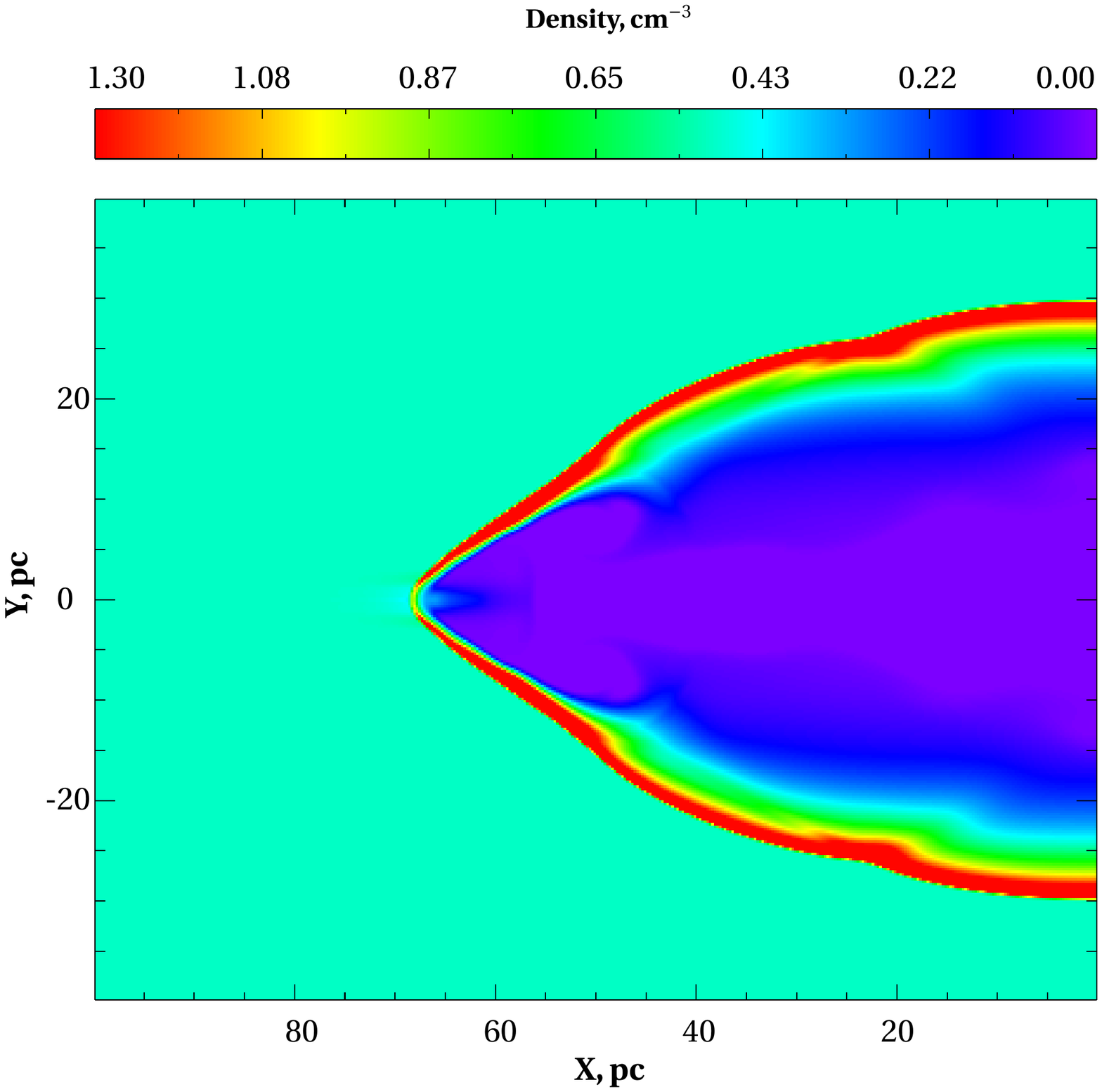}

\caption{Simulation of a transient neutral beam in the ISM. A powerful $L_J=10^{40}\,{\rm erg\,s^{-1}}$ and narrow $r_J=1$~pc beam was operating for 3~kyr in the ISM with the number density $n_{\rm ISM}=0.5\,{\rm cm^{-3}}$. The density snapshot shown corresponds to 60~kyr after the onset of beam heating. By this time only an outer shock and a low-density central region can be identified in the density distribution.}
\label{fig:nb_short}
\end{figure}

Finally, the simulation shown in Fig.~\ref{fig:nb_short} illustrates the evolution of the outer shock long after the end of the NB heating. In this run, a powerful ($L_J=10^{40}\,{\rm erg\,s^{-1}}$) and narrow ($r_J=1$~pc) NB was operating for $3\,10^3$~yr in the ISM with $n_{ISM}=0.5\,{\rm cm^{-3}}$. The snapshot shown corresponds to $60\,10^{4}$~yr. By that time only the outer shock and the hot low-density inner cavity remain visible.  Since both the NB and HJ scenarios possess an outer shock, it might be difficult to differentiate between these scenarios at this evolutionary phase.

\section{Discussion}

Above we demonstrated that a narrow beam of fast neutral hydrogen atoms can produce a wide variety of structures in the ISM. Some of these structures, e.g., an outer shock wave and an underdense central region are reminiscent of similar structures produced by canonical hydrodynamic jets. The problem can be formulated in terms of the energy and momentum injection rate, and the coupling of the flow to the ambient gas. It bears much similarity to generic models of double radio sources \citep[e.g.][]{1973MNRAS.164..243L,1974MNRAS.166..513S} or radio-mode AGN feedback in galaxy clusters \citep[e.g.][]{2000A&A...356..788C}. The key feature of the NB model, is that the injected momentum is low and the propagation velocity is set by the ionization rate. This implies that markedly different parameters of the ISM and of the beam/jet are needed to produce similarly-looking nebulae. Therefore, if we know some of the parameters of the problem, e.g., ISM density, then it becomes much easier to differentiate between the NB and HJ models. In particular, the propagation velocity of NB is a factor of few higher than that of HJ for a plausible range of parameters.  Also, as illustrated in the previous section (see Fig.~\ref{fig:nb_rj}), a hot "tube" and a bright "core" could form naturally in the NB scenario.

Another level of complexity could be added by the time variability (in terms of power and/or direction) of the beam and the non-uniformity of the medium. Of particular interest in application to SS 433, for instance, should be a full 3D picture arising due to precession of the pair of neutral beams. We deffer the detailed discussion of all these possible scenarios for the future work.

Noteworthy, the NB scenario implies that strong shocks coexist with the fast non-thermal particles (in particular 30~MeV protons) that can appear (after ionization of neutral atoms) in both upstream and downstream regions. Moreover, a converging shock can form (see  mid-panel in Fig.~\ref{fig:nb_rj}) for a beam with a boxy shape. It is plausible, that in such configuration the efficiency (or the rate) of particle acceleration will be very high. 

There are also several direct observational implications naturally arising in the NB scenario. Namely, the presence of the fast protons in the ISM should lead to a proton-electron bremsstrahlung emission (X-ray emission is produced by electrons moving in the frame of fast protons). While the efficiency of this process is always small compared to the Coulomb energy losses \citep[e.g.][]{2001ApJ...557..560P}, given the plausible range of the beam power, its contribution to the X-ray spectra might be non-negligible. We note in passing, that adiabatic losses (and any other losses) on top of the Coulomb losses would impact the efficiency of this process. Additionally, non-equilibrium ionization might be important in some of the features formed in the NB scenario, especially in a short-lived "dense core".

Finally, we comment on the requirements for the NB scenario to be relevant for real astrophysical setups. Indeed, certain conditions must be satisfied for a beam of particles to remain neutral (and in a kinetic mode) before it enters denser ambient medium. For instance, photoionisation by the UV radiation might ionise atoms as they move, which could be especially important for hyper-accreting objects. On the other hand, thermal instabilities in a rapidly cooling outflow should result in fragmentation of the continuous disperse flow into a large number of tiny clumps. Such picture is indeed inferred from the optical line emission of the baryonic jets in  SS 433 \citep[e.g.][]{2004ASPRv..12....1F}.  The gas number density and column density of such clumps could strongly affect the regime of the ISM/beam interaction. We defer the discussion of all these effects for a future study.

\section{Conclusions}
Baryonic sub-relativistic jets are known to be launched by hyper-accreting compact objects. Such jets should expand, cool and recombine as they propagate away from the source. This opens a possibility to consider a problem of a beam of neutral atoms propagating through the denser ambient medium. It turns out that such a setup, which is reminiscent of the neutral beam heating technique used for plasma heating in major TOKAMAK facilities, might lead to several very interesting signatures. In particular, this scenario implies (i) a very fast heat-driven propagation of the beam through the medium and (ii) potentially very efficient particle acceleration of "pick-up" ions by the shocks generated by the same beam. Rich phenomenology is predicted to arise in this scenario even without invoking time-dependent beam injection and possible complexity of the surrounding plasma. This scenario has a number of direct observational implications, which might be tested by the observational data on sources featuring supercritical regime of accretion, e.g.  ultraluminous X-ray sources, tidal disruption events and, in future, growing supermassive black holes in the early Universe.

\section*{Acknowledgements}
We are grateful to our reviewer for several important and constructive comments and suggestions. EC, IK, and  RS  acknowledge  partial  support  by  the  Russian  Science  Foundation  grant  19-12-00369. 











\bsp	
\label{lastpage}
\end{document}